\documentclass[a4paper, 11pt]{article}
\usepackage[affil-it]{authblk} 
\usepackage{etoolbox}
\usepackage{lmodern}
\usepackage[a4paper]{geometry}
\usepackage[in]{fullpage}
\usepackage[utf8]{inputenc}
\usepackage{amsmath}
\usepackage{amssymb}
\usepackage{cite}

\usepackage{amsfonts}
\usepackage{graphicx}
\usepackage[dvipsnames]{xcolor}
\usepackage[colorlinks]{hyperref}
\hypersetup{linkcolor=NavyBlue,citecolor=NavyBlue,urlcolor=blue}

\usepackage{soul}

\title{Purely non-perturbative AdS vacua and the swampland}

\makeatletter
\patchcmd{\@maketitle}{\LARGE \@title}{\fontsize{16}{19.2}\selectfont\@title}{}{}
\makeatother

\author[1]{Heliudson Bernardo\footnote{Email: \href{mailto:heliudson@hep.physics.mcgill.ca}{heliudson@hep.physics.mcgill.ca}}}

\author[1]{Suddhasattwa Brahma\footnote{Email: \href{mailto:suddhasattwa.brahma@mcgill.ca}{suddhasattwa.brahma@gmail.com}}}

\author[1]{Keshav Dasgupta\footnote{Email: \href{mailto:keshav@hep.physics.mcgill.ca}{keshav@hep.physics.mcgill.ca}}}

\author[2]{Radu Tatar\footnote{Email: \href{mailto:Radu.Tatar@Liverpool.ac.uk}{Radu.Tatar@Liverpool.ac.uk}}}

\affil[1]{Department of Physics, McGill University,\protect\\ Montreal, QC, H3A 2T8, Canada}

\affil[2]{Department of Mathematical Sciences, University of Liverpool,\protect\\ Liverpool, L69 7ZL, United Kingdom}

\date{\vspace{-5ex}}

\begin{document}

\maketitle


\begin{abstract}
In recent times, a considerable effort has been dedicated to identify certain conditions -- the so-called swampland conjectures -- with an eye on identifying effective theories which have no consistent UV-completions in string theory. In this paper, we examine the anti-de Sitter vacua corresponding to solutions which arise from purely non-perturbative contributions to the superpotential and show that these solutions satisfy the (axionic) weak gravity conjecture and the AdS-moduli scale separation conjecture. We also sketch out their advantages over other constructions.
\end{abstract}



\section{Introduction}
\label{sec:intro}

Moduli stabilization is the main issue to overcome when constructing semi-realistic low-energy effective theories from string compactification. The state-of-the-art approach consists of using fluxes to stabilize the axion-dilaton and complex structure (shape) moduli and non-perturbative effects on D7-branes wrapping cycles in the internal manifold or Euclidean D3-branes to stabilize the K\"ahler structure (see \cite{Dasgupta:1999ss, Gukov:1999ya, Giddings:2001yu, Kachru:2003aw, Balasubramanian:2005zx} for some original references and \cite{Grana:2005jc, Blumenhagen:2006ci} for reviews). In such constructions, an anti-de Sitter (AdS) space is the generic 4d geometry obtained, ultimately, due to supersymmetry. 

On the other hand, the swampland program (see \cite{Brennan:2017rbf, Palti:2019pca} and references therein) proposes identifying theories which cannot have UV completions within string theory. Based on explicitly examples from string theory, there are now several conjectures about such effective theories, the weak gravity conjecture (WGC) being one of the most supported ones \cite{ArkaniHamed:2006dz}. There are different incarnations of the WGC (e.g. electric, magnetic, scalar), depending on which fields we are applying it to, but they all share the feature that gravity should be the weakest force in any effective theory. For $p$-form fields with coupling strength $g_p$, it states that there must exist a charged $(p-1)$-brane whose tension is bounded by its charge times $g_p$ (in Planck units) \cite{Heidenreich:2015nta}. The special case $p=0$ corresponds to the \emph{axionic} WGC in which the decay constant of an axion (0-form) is bounded by its coupling to an instanton. 

In string compactification models, such as KKLT \cite{Kachru:2003aw} and LVS \cite{Balasubramanian:2005zx}, the volume of cycles of the internal direction always appear complexified by axions coming from the integration of forms over the cycles and, thus, axions are ubiquitous in these models. Hence, the same stabilization mechanism which gives mass to the volume moduli automatically fixes the the decay constant of such axions. It is then natural to ask whether these axions satisfy the (axionic version of the) WGC. It was shown in \cite{Moritz:2018sui} that there are regions in the parameter space of the Kallosh-Linde (KL) model \cite{Kallosh_2004, BlancoPillado:2005fn} such that the axion decay constant may violate WGC, and it was quickly realized in \cite{Kallosh:2019axr} that this region is not mandatory for the model to work and that the KL model generically satisfies the WGC. For that to follow, the dependence of the stabilized volume on the parameters of the non-perturbative superpotential was crucial since the no-scale K\"ahler potential in these models is such that the decay constant of the axion depends on the volume of the cycle. 

Motivated by questions about the interplay between fluxes and supersymmetry \cite{Sethi:2017phn}, we recently proposed a racetrack-like mechanism for K\"ahler moduli stabilization in models with vanishing flux potential at the minima of the axion-dilaton and complex moduli ($W_0 = 0$). The model has a Minkowski and an AdS minima, or two AdS minima after a small shift in some parameters of the model, similar to the KL model. However, even though it employs the same ingredients as in the racetrack and KKLT scenarios, the $W_0 =0$ model has the distinct feature of having \textit{two maxima}, therefore achieving the decompactification limit for positive values of the potential. Moreover, the stabilized value of the volume moduli is parametrically different from the one in the KL model. Since the decay constant dependence of the parameters in the model is different from the KL model, this last property motivates us to check whether such a model can satisfy the axionic WGC or not. It is shown in section \ref{sec3} that the axion decay constant in the $W_0=0$ models are always sub-Planckian and that the WGC is consistently satisfied.

Another relevant conjecture to our model is the AdS-moduli scale separation conjecture \cite{Gautason:2018gln}, which limits the mass of the lightest stabilized modulus by the AdS scale $-\Lambda$. In \cite{Blumenhagen:2019qcg, Blumenhagen:2019vgj} the KKLT model was shown to satisfy a log-corrected version of this conjecture, and the results were used to motivate log-corrected refinements for the trans-Planckian censorship conjecture (TCC) and the AdS-moduli and AdS-distance conjectures. In section \ref{sec4} we carry out a similar analysis for the KL and $W_0 = 0$ models. We find that the AdS-moduli conjecture is satisfied for both models without the introduction of log corrections. 

The goal of this note is to show that the $W_0 =0$ model satisfies relevant swampland conjectures. The paper is organized as follows. Section \ref{sec2} is a small overview of the construction with an explanation on how it can be potentially embedded in concrete Calabi-Yau orientifold compactifications, sections \ref{sec3} and \ref{sec4} discusses how the axionic WGC and AdS scale separation conjecture are satisfied, respectively, highlighting the similarities and differences between the two cases. We present our conclusions in section \ref{sec5}. We use Planckian units when not stated otherwise.

\section{New non-perturbative AdS backgrounds}\label{sec2}

In this section we review the model introduced in \cite{Bernardo:2020lar} characterised by a potential with two minima, corresponding to a Minkowskian and an AdS vacua. Although this feature is also present in the KL model, the superpotential  considered here,
\begin{equation}
\label{supot}
 W = \sum_{j=1}^3 A_j e^{ia_j\rho}, 
\end{equation}
is different, with zero contribution from the fluxes as explained in section \ref{W0=0} and three non-perturbative contributions, which may correspond to three different stacks of D7-branes. Moreover, even neglecting large volume corrections to the no-scale K\"ahler potential, the final potential goes to zero from above for large values of the 4-cycle modulus, i.e. the potential has two maxima in contrast to the KKLT and KL potentials. 

\begin{figure}[t]
    \centering
    \includegraphics[width=0.7\columnwidth]{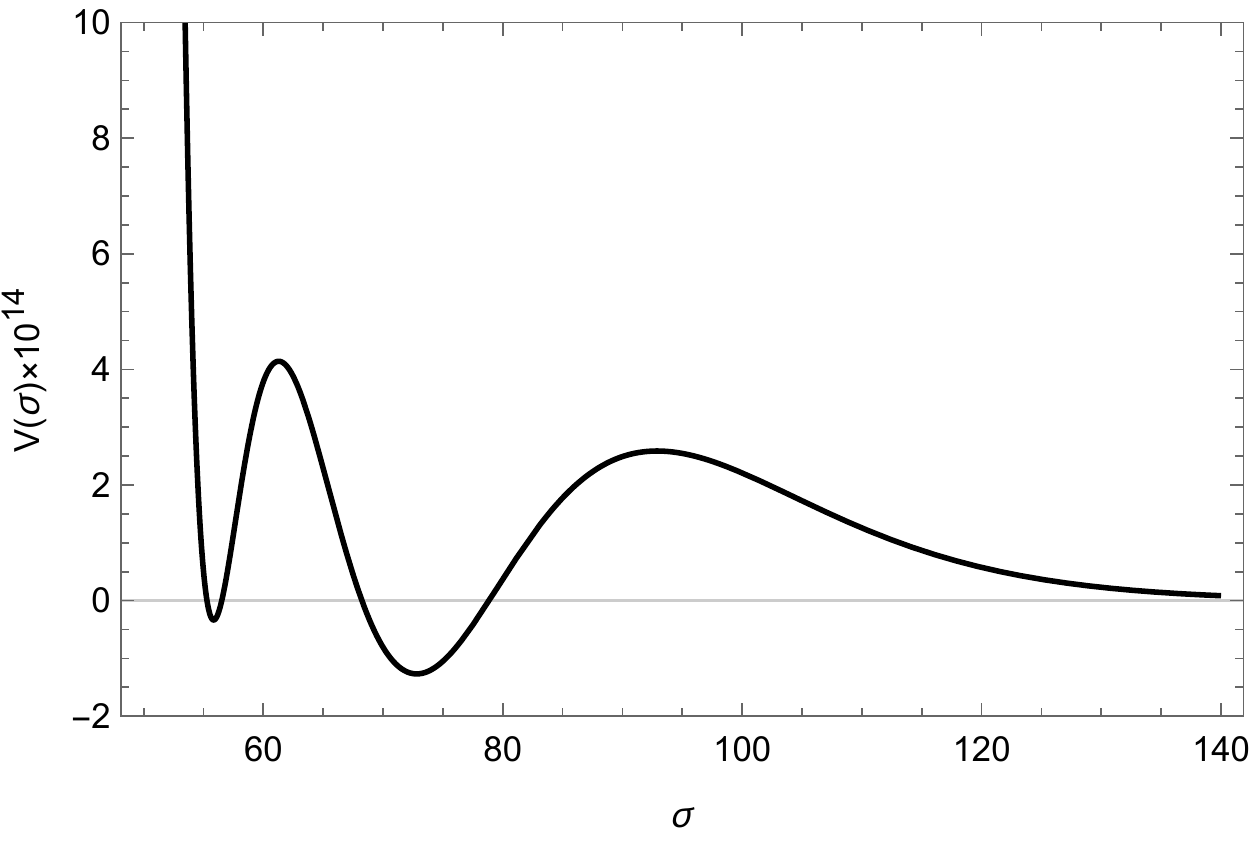}
    \caption{Potential for the $W_0$ model, with parameter values $a_1 = 2\pi/100$, $a_2 = 2\pi/80$, $a_3=2\pi/70$, $A_2 = 0.98$, $A_3 = -1.05$ and $\delta A_1 = 0.007 A_1$. The first AdS minimum is located at $\sigma_0 \approx 54.2$.}
    \label{W=0model}
\end{figure}

The condition for a susy preserving Minkowski vacuum at $\rho = i\sigma_0$, 
\begin{equation}
    \sum_{j=1}^{3} A_j\left(1 +a_j \frac{2\sigma_0}{3}\right)e^{-a_j\sigma_0} = 0 = \sum_{j=1}^3 A_j e^{-a_j \sigma_0},
\end{equation}
fixes one of the parameters of the model, say $A_1$,  and $\sigma_0$ to be
\begin{align}\label{A1parameter}
 A_1 &=- \left(A_2 e^{(a_1 -a_2)\sigma_0} + A_3 e^{(a_3 -a_2)\sigma_0}\right), \\  \label{sigma0}
 \sigma_0 &= \frac{1}{a_3 -a_2}\ln \left(-\frac{a_3 - a_1}{a_2 - a_1}\frac{A_3}{A_2}\right), 
\end{align}
where we are assuming $a_1 \neq a_2 \neq a_3$, as explained in \cite{Bernardo:2020lar}. Assuming that all parameters are real, from the expression for $\sigma_0$ we should have $a_3> a_2$ and there are two possibilities for the hierarchy between the other $a$'s:
if $\text{sign}(A_3) = - \text{sign}(A_2)$, then $a_3> a_1$ and $a_2>a_1$; while for $\text{sign}(A_3) = \text{sign}(A_2)$ we should have $a_2<a_1$ and $a_3> a_1$. Thus, either $a_3>a_2>a_1$ or $a_3> a_1> a_2$ in order for $\sigma_0>0$. The potential has also an AdS minimum at $\sigma_1>\sigma_0$, see Figure \ref{W=0model}. 

Similarly to the KL model, by changing $A_1$ away from the fixed value (\ref{A1parameter}), we can promote the first minimum to be also AdS. If the change $\delta A_1$ is small, the position of the minimum is practically unchanged and
\begin{equation}
    V(\sigma_0) \approx -\frac{3}{8\sigma_0^3}\delta A_1^2e^{-2a_1\sigma_0}.
\end{equation}
Furthermore, given that the minimum was initially Minkowski, we can use the same arguments from \cite{BlancoPillado:2005fn} to argue for the positivity of the mass matrix of the stabilized dilaton and complex moduli. This guarantees the absence of tachyons and a parametrically smaller susy breaking scale (as compared to the moduli mass scales) after uplifting this shallow vacuum to dS. In the next subsection we discuss the life-time of the uplifted vacuum.

\subsection{Uplift and dS lifetime}
Although, in this paper, we shall mainly discuss different aspects of the AdS vacua described above, let us give a quick review of the dS vacua one gets by uplifting them, as shown in \cite{Bernardo:2020lar}. The main characteristic of these solutions is that they present a lifetime much smaller than the standard KKLT and LVS scenarios. The uplifting to dS\footnote{We are assuming an uplifitng mechanism similar to KKLT here. However, note that difficulties with respect to this uplifting procedure has been recently discussed in \cite{Bernardo:2020lar}.} for the shallower of the two minima can be taken care of by adding an anti-D3 brane, as was done for the original KL model. The resulting potential has a dS minima as well as an AdS one (in the case of the minimum three non-perturbative terms). The decay channel for the resulting dS vacuum is to the AdS minima and this gives the leading order contribution to the decay time. As has been discussed in \cite{BlancoPillado:2005fn}, the decay time for such a configuration can be calculated, provided we i) assume the thin-wall approximation, ii) assume the tension of the bubble wall remains unaffected by the uplift an iii) assume that the depth of the AdS minima remains the same in spite of the uplifting since the shallower of the two minima was very close to Minkowski before the uplift. On assuming these well-known conditions, the lifetime can be approximated as
\begin{eqnarray}
 \tau_{\rm dS} = e^{\frac{24 \pi^2}{|V_0|}\,\, \frac{(C-1)^2)}{C^2 (2C-1)^2}}\,,
\end{eqnarray}
where $V_0$ is the depth of the shallower of the two minima (before uplift) and $V_1$ is the deeper of the two AdS minima and $C = \sqrt{|V_1|/|V_0|}$. As has been shown in \cite{Bernardo:2020lar}, the lifetime, although much shorter than other standard proposals, cannot be made to be short enough to be compatible with the recently proposed trans-Planckian censorship conjecture (TCC) \cite{Bedroya:2019snp,Bedroya:2019tba} by tuning the parameters of this model, as long as one still obeys the supergravity approximation.

\subsection{Models with vanishing flux superpotential}\label{W0=0}

The original racetrack scenario used the sum of two exponential terms $A e^{ia \tau} + B e^{ ib \tau}$ to fix the dilaton field at weak coupling in heterotic strings \cite{Krasnikov:1987jj,Taylor:1990wr} (see also \cite{Dine:1999dx,Dine:1985rz}). For $a = 2\pi/N$, $b = 2\pi/M$, the superpotential for the axion-dilaton field $\tau$ can be imagined to be generated by gaugino 
condensates for the group $SU(N) \times SU(M)$. When $N$ and $M$ are large and close to each other, the supersymmetric minimum occurs at
\begin{equation}
 \tau = \frac{N M}{M-N} \mbox{log}\left(- \frac{M B}{N A}\right), 
\end{equation}
whose imaginary part is very large. Therefore, $e^{-{\rm Im} \tau}$ is small and so the dilaton is stabilized at the weak coupling regime. 

In \cite{Escoda:2003fa}, the authors adapted the racetrack idea to KKLT type models (in which the relevant moduli are the complex K\"ahler moduli) by adding an extra contribution $W_0$ to the superpotential and the idea was extended to the KL model in \cite{Kallosh_2004}. In our discussion we want to preserve $W_0 = 0$ as in the original racetrack. The relevant geometries that are used to obtain $W_0 = 0$  were discussed in \cite{Denef:2004dm} (and further clarified in \cite{Curio:2006ea}). In these papers, there are two different constructions that are relevant to us. The first refers to IIB  compactified on a Calabi-Yau 3-fold $Z$ with a holomorphic involution $\hat{\Omega}$ with fixed loci corresponding to O3 and O7 planes. Moreover, $Z$ has arbitrary $h^{2,1}$ three cycles and $h^{1,1}$ cycles and D branes are introduced to cancel the RR tadpoles of the orientifold planes. The second construction considers an F theory compactification on a Calabi-Yau fourfold $X$ which is an elliptic fibration $\pi$ over a threefold $B$. Type IIB is  compactified on $B$ and D7 branes are introduced at singularities of $\pi$. The complex combination between axion and dilaton  at a point on $B$ is the fiber $\pi^{-1}$ over $B$ at that point. If the 
singularities of $\pi$ are $D_4$ singularities, four coincident D7 branes and an O7 plane are introduced and we recover $Z$ as a double cover of $B$. 

The sizes of the $h^{2,1}$ three cycles are fixed by RR and NS fluxes and the sizes of the $h^{1,1}$ four cycles are fixed by D7 branes wrapped on them. The fixing of the sizes of the $h^{1,1}$ four cycles is done by a superpotential generated on the D7 branes. This superpotential has been argued to be nonzero only if the arithmetic genus $\sum_{j=0}^{3} h^{0,j}$ is one and \cite{Denef:2004dm} searched for models with this property among divisors of Calabi-Yau 4-folds $X$. Using the relation between $X$, its base $B$ and the double cover of its base $Z$, this led to a search for a class of constructions for $B$ which are $P^1$ bundles over toric surfaces and their orientifold limits $Z$ which are elliptic fibres over $P^2$. One such construction has $h^{1,1} = 2$ and $h^{2,1} = 272$ but, due to an $Z_6 \times Z_{18}$ action on the moduli space, only two complex deformations $z_1$ and $z_2$ are independent. By fixing them at $z_1 = z_2 = i$ and the dilaton, one finds a quantized vacuum 
with $W_0=0$ which is exactly what we want before introducing nonperturbative terms. The nonperturbative terms appear as $\sum_{j=1}^2 b_j e^{ 2 \pi i a_j \rho_j}$ where $\rho_1, \rho_2$ are the complexified volumes of the 2 independent 4-cycles implied by $h^{1,1} = 2$. 

This immediately raises the question of how to deal with more than one K\"ahler parameter, i.e. $h^{1,1}>1$. In our discussion, we are wrapping branes on one cycle whose size becomes fixed but it is important to ask what about the sizes of the unwrapped cycles. As we discuss Minkowski vacua, we can apply the discussion of \cite{BlancoPillado:2005fn} and \cite{Krefl:2006vu} in this context. Collectively denoting the complex moduli  $(x_1, \cdots, x_m)$ by $x$, the K\"ahler moduli $(\rho_1, \cdots, \rho_n)$ by $\rho$ and the dilaton-axion by $\tau$, the supersymmetry condition for vanishing cosmological constant $W(\rho^0, \tau^0, x^0) = 0 $ is $\partial_I W = 0$, where $I$ runs through all moduli.

If we allow a non-perturbative racetrack type potential for each Kahler modulus $\rho_i$
\begin{equation}
W_{np} = \sum_j (C_{j} e^{ia_j  \rho_j} - D_j e^{ib_j \rho_j})
\end{equation}
and write the flux superpotential as $A(x_0) + \tau B(x_0)$. 
The Minkowski vacua occur at 
\begin{equation}
B(x^0) = 0,\quad A(x^0) + \sum_j (C_{j} e^{i a_j \rho_j^0} - D_j e^{ib_j \rho_j^0}) =0.
\end{equation}
This can be interpreted as choosing the fluxes such that the complex moduli are stabilized at $(x_1^0, \cdots, x_m^0)$ and then fixing 
$\rho$ as function of $x^0$. 

In our case, let us consider $n = h^{1,1} - 1$ so the K\"ahler moduli, fixed by the choice of complex moduli, are $\rho_1, \cdots, \rho_{h^{1,1} -1}$, this comes just from requiring a Minkowski vacuum for the potential. Then we have a superpotential for $\rho_{h^{1,1}}$:
\begin{equation}
W = \sum_{j=1}^{3} A_{h^{1,1}, j} e^{i a_{h^{1,1},j} \rho}    
\end{equation}
which leads to the Minkowski and AdS minima appearing in our discussion. This is same as equation (\ref{supot}) when $a_j$ are relabelled as $a_{h^{1,1}, j}$. Therefore our model is valid for any value of $h^{1,1}$ with the understanding that $h^{1,1} - 1$ number of K\"ahler moduli are fixed by the choice of complex moduli in a Minkowski vacuum. Models with Minkowski vacuum $W_0 = 0$ were also recently considered in \cite{Kanno:2017nub} and \cite{Kanno:2020kxr} by pursuing the ideas originally proposed in \cite{DeWolfe:2004ns} to restrict the set of complex deformations to an algebraic set which satisfies $W_0 = 0$. In particular, \cite{Kanno:2020kxr} consider an F-theory compactification on $K3 \times K3$ which implies a pretty large value for $h^{1,1}$, it would be interesting to see how to explicitly fix all but one of the K\"ahler deformations as a function of complex deformation and follow up with our superpotential (\ref{supot}).

\section{Axionic WGC conjecture}\label{sec3}

The weak gravity conjecture \cite{ArkaniHamed:2006dz} for axions states that an axion ($0$-form) should couple with an instanton (($-1$)-brane) such that  (see \cite{Rudelius:2014wla, Rudelius:2015xta, Heidenreich:2015wga,Brown:2015iha,Bachlechner:2015qja,Junghans:2015hba} for discussions on the axionic WGC and \cite{Palti:2019pca} for a review)
\begin{equation}
    f S_{E}\leq M_{\text{Pl}},
\end{equation}
where $f$ is the axion decay constant and $S_{E}$ is the Euclidean action for the instanton. If on top of that we want to have a controlled instanton expansion and the contribution of the harmonic term induced by this instanton to the potential is non-negligible, then the axionic WGC implies that $f$ should be sub-Planckian. As in \cite{Moritz:2018sui} we will assume that a strong form of the conjecture: it should apply for any harmonic, regardless of the non-perturbative effect that generated it.

Recalling that the 4-cycle volume modulus $\sigma$ is always accompanied by an axionic field $\phi$, both appearing in the scalar component of the chiral multiplet as $\rho = \phi +i\sigma$, in \cite{Moritz:2018sui} it was pointed out that the choice of parameter for stabilizing $\sigma$ could be incompatible with the strong form of the weak gravity conjecture applied for $\phi$, which would require its decay constant to be sub-Planckian. However, in \cite{Kallosh:2019axr} it was shown that the KL model can indeed satisfy this form of WGC with a choice of parameters entirely compatible with the stabilization mechanism, although the authors explicitly criticize the specific form of the conjecture. In the following, we study whether we can also satisfy the WGC in our model, i.e., check if we can consistently select parameters such that the axion has a sub-Planckian decay constant. 

In order to calculate the axion decay constant, we will start from the full potential and write $\rho = \phi + i\sigma$, assuming only $a_i$ to be real. For
\begin{equation}
    W = W_0 + \sum_j A_j e^{ia_j\rho}, \quad K = -3\ln(-i(\rho -\Bar{\rho})),
\end{equation}
we have
\begin{align}\label{generalpotential}
    V(\phi, \sigma) &= \frac{1}{2\sigma^2}\sum_j^{N} |A_j W_0| a_j e^{-a_j\sigma} \cos{(a_j\phi +\alpha_j)} + \frac{1}{6\sigma}\sum_j^{N} |A_j|^2\left(a_j^2 +\frac{3}{\sigma}a_j\right)e^{-2a_j\sigma} +\nonumber\\
    &+\frac{1}{3\sigma}\sum_{j<k}^{N}|A_jA_k|\left(a_ja_k +\frac{3}{2\sigma}(a_j +a_k)\right)e^{(a_j +a_k)\sigma}\cos{((a_j - a_k)\phi +\alpha_{jk})},
\end{align}
where $\alpha_j =\arg(A_j\overline{W}_0)$ and $\alpha_{jk}= \arg(A_j\overline{A}_k)$. These phase factors are set to zero in the following, since it is sufficient to consider real values of $A_i$. For $N=2$, we get the full potential of the KL case studied in \cite{Moritz:2018sui}. Note that in general for $N$ non-perturbative terms in the superpotential, we could have up to $N + N(N-1)/2$ harmonic terms in the final potential. Of course, we are interested in the case $W_0=0$ and $N=3$, and thus we have three possible harmonics, denoted by $(i,j)$ (with $i>j$) corresponding to the amplitude proportional to $A_i A_j$.

The kinetic term of $\phi$ can be read from the K\"ahler metric which implies that at a minimum with $\sigma = \sigma_0$, the canonically normalized field corresponding to $\phi$ will be
\begin{equation}
    \Tilde{\phi} = \sqrt{\frac{3}{2}}\frac{\phi}{\sigma_0}.
\end{equation}
Thus, from the general potential (\ref{generalpotential}) for $W_0=0$, the possible axion decay constants are
\begin{equation}
    f_{ij} = \sqrt{\frac{3}{2}}\frac{1}{(a_i-a_j)}\frac{1}{\sigma_0},
\end{equation}
and in order for the WGC to be satisfied we need $f_{ij}<1$ (we're using natural units) for the dominant $(i,j)$ harmonic.

We see that \emph{a priori} we can have $a_i\sigma_0>1$ but $(a_i - a_j)\sigma_0 <1$ such that $f_{ij} >1$. So, we should look for a region in the parameter space where this does not happen, that is, we need to find choices of parameters such that $(a_i-a_j)\sigma_0>1$. Fortunately, the harmonic $(3,2)$ can easily satisfy WGC if the log factor in (\ref{sigma0}) is greater than $\sqrt{3/2}$, i.e. if we choose
\begin{equation}
    A_3 > e^{\sqrt{3/2}}|A_2|\frac{a_2 - a_1}{a_3 - a_1} \implies \sqrt{\frac{3}{2}}\ln \left(-\frac{a_3 - a_1}{a_2 - a_1}\frac{A_3}{A_2}\right) >1
\end{equation}
Furthermore assuming $a_3> a_2> a_1$ we have
\begin{equation}
    f_{31} = \sqrt{\frac{3}{2}}\frac{1}{a_3-a_1}\frac{1}{\sigma_0}< \frac{a_3 -a_2}{a_3-a_1} < 1,
\end{equation}
and the harmonic $(3,1)$ automatically yields a sub-Planckian decay constant. If we further assume that $a_3 -a_2< a_2- a_1$ the decay constant $f_{21}$ will satisfy
\begin{equation}
    f_{21} = \sqrt{\frac{3}{2}}\frac{1}{a_2-a_1}\frac{1}{\sigma_0}< \frac{a_3 -a_2}{a_2-a_1} < 1,
\end{equation}
and then all possible decay constants are sub-Planckian, regardless of which terms dominate the potential. We see that there is a region in the parameter space where the WGC can be satisfied without invalidating the instanton expansion and the stabilization of $\sigma$ at large values. 

\section{AdS-moduli scale separation conjecture}\label{sec4}

Recently, it was shown in \cite{Blumenhagen:2019qcg, Blumenhagen:2019vgj} that the KKLT AdS minimum satisfies a log-corrected version of the AdS-moduli scale separation conjecture \cite{Gautason:2018gln}. This conjecture states that there should not be a scale separation between the mass $m$ of the lightest stabilized modulus and the AdS scale $|\Lambda|$,
\begin{equation}
    m \leq c |\Lambda|^{1/2},
\end{equation}
where $c$ is a positive $\mathcal{O}(1)$ number. Another relevant conjecture on AdS vacua is the AdS distance conjecture, introduced in \cite{Lust:2019zwm}, which states that in the limit $\Lambda \to 0$ there exists a tower of light states with mass satisfying
\begin{equation}
    m_{\text{tower}} = c'|\Lambda|^{\alpha},
\end{equation}
with $\alpha>0$. A stronger version fixes $\alpha =1/2$ for supersymmetric AdS, corresponding to no scale separation between the mass of the states and the AdS curvature scale.

Given the difference on how the AdS vacuum for the KL and the $W_0 =0$ model are constructed, in this section, we investigate whether the scale separation conjecture is satisfied or not for those models. The analysis will be sensible when 4-cycle moduli mass is the relevant mass scale for the AdS scale separation conjecture, which is the case for KKLT \cite{Blumenhagen:2020dea}. Given that there is no modification to the complex structure moduli stabilization mechanism, we expect that to be also the case in the models we are interested in.

After turning off the axion ($\phi =0$) and considering the parameters in the superpotential to be real, the potential in both cases can be read from
\begin{equation}\label{reducedpotential}
    V(\sigma) = \frac{1}{6\sigma}\sum_{i,j}^{N}A_i A_j \left(a_i a_j + \frac{3}{\sigma}a_j\right)e^{-\sigma(a_i + a_j)} + \frac{W_0}{2\sigma^2}\sum_i^N A_i a_i e^{-a_i \sigma},
\end{equation}
where $N=2$ for the KL model and $W_0 = 0$ and $N=3$ for our model. Its first and second derivatives are given by
\begin{align}
    V'(\sigma) &= -\sum_{i,j}A_iA_j \left(\frac{2}{3}\frac{a_i a_j}{\sigma^2}+ \frac{a_j}{\sigma^3} + \frac{1}{2}\frac{a_j^2}{\sigma^2}+ \frac{1}{3}\frac{a_i a_j^2}{\sigma}\right)e^{-(a_i+a_j)\sigma} - \frac{W_0}{2\sigma^2}\sum_i A_i\left(a_i^2+2\frac{a_i}{\sigma}\right)e^{-a_i \sigma},\\
    V''(\sigma) &= \sum_{i,j}A_iA_j \left(\frac{7}{3}\frac{a_i a_j}{\sigma^3}+ 2\frac{a_j^2}{\sigma^3}+ \frac{15}{6}\frac{a_ia_j^2}{\sigma^2}+\frac{1}{2\sigma^2}a_j^3+ 3\frac{a_j}{\sigma^4}+\frac{1}{3}\frac{a_i^2a_j^2}{\sigma}+ \frac{1}{3}\frac{a_ia_j^3}{\sigma}\right)e^{-(a_i+a_j)\sigma} +\nonumber\\
    &+ W_0\sum_i A_i \left(2\frac{a_i^2}{\sigma^3}+3\frac{a_i}{\sigma^4}+\frac{1}{2}\frac{a_i^3}{\sigma^2}\right)e^{-a_i\sigma}.
\end{align}
Now let us impose the condition for a supersymmetric vacuum at $\sigma = \sigma_0$, i.e. $DW(\sigma_0) = 0$. For the KL model, this condition reduces to
\begin{equation}
    W_0 = -\sum_j A_j\left(1+ \frac{2}{3}a_j\sigma_0\right)e^{-a_i \sigma_0},
\end{equation}
while for our $W_0 =0$ model, it is simply
\begin{equation}
    \sum_j A_j e^{-a_j\sigma_0} = - \frac{2}{3}\sigma_0\sum_j A_j a_je^{-a_i\sigma_0}.
\end{equation}
For both models, imposing the susy vacuum at $\sigma_0$ results into
\begin{align}
    V(\sigma_0) = -\frac{1}{6\sigma_0}\left(\sum_i A_i a_i e^{-a_i\sigma_0}\right)^2, \qquad  V'(\sigma_0) = 0,
\end{align}
and
\begin{align}
    V''(\sigma_0) &= \frac{1}{3\sigma_0^3}\left(\sum_i A_i a_i e^{-a_i\sigma_0}\right)^2 +\frac{7}{6\sigma_0^2}\left(\sum_i A_i a_i e^{-a_i\sigma_0}\right)\left(\sum_jA_j a_j^2e^{-a_j\sigma_0}\right) +\nonumber\\
    &+\frac{1}{3\sigma_0}\left(\sum_i A_i a_i^2e^{-a_i\sigma_0}\right)^2.
\end{align}

Thus, the mass of the 4-cycle volume can be obtained by evaluating the expression for $V''(\sigma_0)$ with the parameter values chosen to get the Minkowski or AdS minimum. Before doing that, we need to recall that $\sigma$ is not canonically normalized. In fact, from the K\"ahler metric, the canonically normalized field is $\phi = \sqrt{3/2}\ln \sigma$, such that its mass is given by
\begin{equation}
    m^2_{\phi} = \frac{2}{3}\sigma_0^{2}V''(\sigma_0).    
\end{equation}

Denoting $\Lambda = V(\sigma_0)<0$, we can write
\begin{equation}\label{phimass}
     m^2_{\phi} \equiv \left(m_{\phi, \text{M}} +\frac{7}{\sqrt{12}}\sqrt{-\Lambda}\right)^2 + \frac{11}{4}\Lambda,
\end{equation}
where we have defined
\begin{equation}\label{Minkmassscale}
    m_{\phi,\text{M}}^2 \equiv \frac{2\sigma_0}{9}\left(\sum_i A_i a_i^2e^{-a_i\sigma_0}\right)^2.
\end{equation}
Note that although it has the same structure of the mass when the vacuum is Minkowski, for $\Lambda \neq 0 $ its value is not the same as $m^2_{\phi}$ in the Minkowski vacuum, since not only the parameters change between the two cases, but also the location of the minimum ($\sigma_0$) is different. But assuming that the difference between the parameters in the Minkowski and AdS cases to be small (which would require a small $\Lambda$ for consistency), we can approximate $m_{\phi, \text{M}}$ to be the mass at the Minkowski minimum. The mass $m_\phi^2$ of the canonically normalized volume above has a dependence on $\Lambda$ which is parametrically different from the corresponding expression for the KKLT case \cite{Blumenhagen:2019vgj},
\begin{equation}
    m^2_{\phi, \rm KKLT} = -\Lambda (2+ 5a \sigma_0 + 2a^2\sigma_0^2), \quad \Lambda = -\frac{A^2 a^2}{6\sigma_0}e^{-2 a\sigma_0},
\end{equation}
and so the volume mass for the KL and $W_0 =0$ models will have different behaviour in the limit $\Lambda \to 0$ as
compared to the KKLT model.

Up to now, there were no approximations invoked in order to get the results above. To check the relation with the AdS scale separation conjecture explicitly, let us compute the behaviour of $m^2_{\phi}$ for small $\Lambda$, so that can use analytic expressions (for the first minimum) to relate the value of $\sigma_0$ to $\Lambda$. Furthermore, note that for the limit $\Lambda \to 0$ does not necessarily imply that we are back to the prior Minskowski vacuum; this will only be so if we take this limit while keeping $\sigma_0$ fixed. 

We need to determine how $m^2_{\phi, \text{M}}$ depends on $\Lambda$ to check whether there is scale separation or not. For the KL model, we have
\begin{equation}
    \Lambda \approx -\frac{3 \delta W_0^2}{8\sigma_0^3},
\end{equation}
that can be inverted to give
\begin{equation}
    \sigma_0 \approx \left(-\frac{8\Lambda}{3\delta W_0^2}\right)^{-1/3}.
\end{equation}

For the model with $W_0=0$ we have
\begin{equation}
    \Lambda \approx -\frac{3}{8\sigma_0^3}\delta A_1^2e^{-2a_1\sigma_0},
\end{equation}with 
and since we are interested in the $\Lambda \to 0$ limit, we can take 
\begin{equation}
    \sigma_0 \simeq -\frac{\ln(-\Lambda)}{2a_1}.
\end{equation}

We explicitly see that for both models the $\Lambda \to 0$ limit corresponds to the infinite distance limit in field space, as required by the AdS distance conjecture. In this case, the mass scale (\ref{Minkmassscale}) is exponentially suppressed and the mass of the moduli limits to
\begin{equation}
    m^2_{\phi} \to \frac{4}{3}|\Lambda|.
\end{equation}
Thus, for both models the $\Lambda$ dependence in $m^2_{\phi, \text{M}}$ does not dominate $m^2_{\phi}$ and the scale separation conjecture is satisfied without logarithmic corrections, in contrast to the KKLT case. 


\section{Conclusion}\label{sec5}

An essential feature of the KKLT construction which have been recently questioned, even before considering issues related to uplifting, is the fine-tuned, small value of the Gukov-Vafa-Witten superpotential in the vacuum, denoted by $W_0 \sim e^{\mathcal{K}/2} \left\langle \int G \wedge \Omega \right\rangle$, over some Calabi-Yau orientifold. The condition $\left\langle D_iW \right \rangle= 0,\, \left\langle W \right \rangle=0$ is needed for having a supersymmetry preserving Minkowski vacuum, and therefore, choosing a value of $W_0\neq 0$ implies that the vacuum is not supersymmetric anymore. In \cite{Sethi:2017phn}, it was argued that the validity of the perturbative solution is questionable when one considers a non-zero $W_0$. This led us to consider $4$-d EFT models of dS vacua which have $W_0=0$ as the starting point in\cite{Bernardo:2020lar}, and we realized that such models typically have lifetimes much smaller than KKLT. In this work, we examine the AdS vacua corresponding to such models in which we tune the fluxes such that once the complex structure moduli are stabilized, we have $W_0=0$. We show that such models are possible to construct by sketching out how to deal with more than one K\"ahler moduli, as is often required for the relevant geometries when dealing with a vanishing flux superpotential.

The main result of this paper was to show that the new class of AdS vacua, which are constructed purely from non-perturbative terms in the superpotential, satisfies the axionic WGC and the no-scale separation conjecture. Apart from usual swampland considerations, there is a specific objective behind checking these conjectures for our model. It was argued in \cite{Moritz:2017xto} that the backreaction of the uplift, in the effective 4-d model of KKLT, is large enough to flatten the potential, thereby preventing the formation of the dS minimum. Although this was shown to be not true in the relevant parameter space of KKLT \cite{Kallosh:2019axr}, the consensus agreement which came out of this discussion was that the racetrack variety of KKLT (specifically, the KL model) does not have any such problems of backreaction due to the uplift. However, this led the authors of \cite{Moritz:2018sui} to posit that although the racetrack fine-tuning can avoid these flattening effects, they necessarily violate the (axionic) WGC. Once again, this claim was contested for the relevant parameter space of the model \cite{Kallosh:2019axr}. However, these arguments make it clear that our model, with solely non-perturbative terms in the superpotential, needed to be checked for its consistency with the axionic WGC (since we already know that the backreaction flattening would be automatically absent for this mechanism). This led us to explicitly show that the decay constants derived from the potential are always sub-Planckian in this case, in the region of parameter space which is of interest. Interestingly, we also find that the stronger version of the  AdS-scale separation conjecture is obeyed by this class of vacua, as opposed to KKLT, which requires an additional logarithmic correction to be incorporated. Thus, unlike in KKLT, there is no log-type scale separation for these models. In the future, our main goal would be to construct an explicit realization of this solution with a vanishing flux superpotential.

\section*{Acknowledgments}
The research of HB and KD is supported by funds from NSERC. SB is supported in part by the NSERC (funding reference CITA 490888-16) through a CITA National Fellowship and by a McGill Space Institute fellowship.




\bibliographystyle{bibstyle} 
\bibliography{References}





\end{document}